\documentclass[conference]{IEEEtran}
\IEEEoverridecommandlockouts
\usepackage{cite}

\usepackage{algorithm}
\usepackage{algorithmic}
\usepackage{graphicx}
\usepackage{textcomp}
\usepackage{xcolor}
\usepackage{listings}
\usepackage{courier}
\usepackage{subcaption}
\usepackage{amsmath, amssymb, amsthm, bm}
\usepackage{booktabs}
\usepackage{longtable}
\usepackage{multirow}
\usepackage{dsfont}

\newcommand*{\thead}[1]{\multicolumn{1}{c}{\bfseries #1}}

\def\BibTeX{{\rm B\kern-.05em{\sc i\kern-.025em b}\kern-.08em
    T\kern-.1667em\lower.7ex\hbox{E}\kern-.125emX}}

\begin{document}

\title{Hierarchical Graph Matching Network for Graph Similarity Computation}

%\author{
%	Xiao Yan,\ \ Jinfeng Li,\ \ Xinyan Dai,\ \ Hongzhi Chen,\ \ James Cheng\\
%	Department of Computer Science\\
%	The Chinese University of Hong Kong\\
%	Shatin, Hong Kong \\
%	\texttt{\{xyan, jfli, xydai, hzchen, jcheng\}@cse.cuhk.edu.hk} \\
%	%% examples of more authors
%	%% \And
%	%% Coauthor \\
%	%% Affiliation \\
%	%% Address \\
%	%% \texttt{email} \\
%	%% \AND
%	%% Coauthor \\
%	%% Affiliation \\
%	%% Address \\
%	%% \texttt{email} \\
%	%% \And
%	%% Coauthor \\
%	%% Affiliation \\
%	%% Address \\
%	%% \texttt{email} \\
%	%% \And
%	%% Coauthor \\
%	%% Affiliation \\
%	%% Address \\
%	%% \texttt{email} \\
%}

%\author{\IEEEauthorblockN{*Haibo Xiu,\ \ Xiao Yan,\ \ *Xiaoqiang Wang,\ \ James Cheng,\ \ Lei Cao}
%\IEEEauthorblockA{\textit{Department of Computer Science} \\
%\textit{The Chinese University of Hong Kong}\\
%Shatin, Hong Kong \\
%\{sydeng7, xyan, kwng6, cyjiang7, jcheng\}@cse.cuhk.edu.hk}
%}

\author{\IEEEauthorblockN{Haibo Xiu}
\IEEEauthorblockA{\textit{College of Computer Sci. \& Tech.} \\
\textit{Zhejiang University}\\
xiu@zju.edu.cn}
\and
\IEEEauthorblockN{Xiao Yan}
\IEEEauthorblockA{\textit{Dept. of Computer Sci. \& Eng.} \\
\textit{The Chinese University of Hong Kong}\\
xyan@cse.cuhk.edu.hk}
\and
\IEEEauthorblockN{Xiaoqiang Wang}
\IEEEauthorblockA{\textit{College of Computer Sci. \& Tech.} \\
\textit{Zhejiang University}\\
xq.wang@zju.edu.cn}
\and
\IEEEauthorblockN{James Cheng}
\IEEEauthorblockA{\textit{Dept. of Computer Sci. \& Eng.} \\
\textit{The Chinese University of Hong Kong}\\
jcheng@cse.cuhk.edu.hk}
\and
\IEEEauthorblockN{Lei Cao}
\IEEEauthorblockA{\textit{CSAIL} \\
\textit{Massachusetts Institute of Technology}\\
lcao@csail.mit.edu}
%\and
%\IEEEauthorblockA{\textit{Department of Computer Science} \\
}

\maketitle

\begin{abstract}
	Graph edit distance / similarity is widely used in many tasks, such as graph similarity search, binary function analysis, and graph clustering. However, computing the exact graph edit distance (GED) or maximum common subgraph (MCS) between two graphs is known to be NP-hard. In this paper, we propose the hierarchical graph matching network (HGMN), which learns to compute graph similarity from data. HGMN is motivated by the observation that two similar graphs should also be similar when they are compressed into more compact graphs. HGMN utilizes multiple stages of hierarchical clustering to organize a graph into successively more compact graphs. At each stage, the earth mover distance (EMD) is adopted to obtain a one-to-one mapping between the nodes in two graphs (on which graph similarity is to be computed), and a correlation matrix is also derived from the embeddings of the nodes in the two graphs. The correlation matrices from all stages are used as input for a convolutional neural network (CNN), which is trained to predict graph similarity by minimizing the mean squared error (MSE). Experimental evaluation on 4 datasets in different domains and 4 performance metrics shows that HGMN consistently outperforms existing baselines in the accuracy of graph similarity approximation.     
\end{abstract}


\section{Introduction}\label{sec:intro}

Graph is a powerful format of data representation and is widely used in areas such as social networks~\cite{xiang2010modeling, tsai1979error, konstas2009social}, biomedical analysis~\cite{borgwardt2005protein, eckert2007molecular}, recommender systems~\cite{debnath2008feature}, and computer security~\cite{shang2010detecting, kinable2011malware}. Graph distance (or similarity)~\footnote{For conciseness, we refer to both graph distance and graph similarity as graph similarity as it is easy to transform a distance measure into a similarity measure.} is important for many graph-based tasks such as graph similarity search~\cite{zeng2009comparing,zager2008graph}, binary function analysis~\cite{xu2017neural} and anomaly detection~\cite{papadimitriou2010web}. For example, in binary function analysis, there is a database of control-flow graphs that are known to have problems, and the goal is to find if a software is prone to these problems. A natural solution is to search in the graph database to decide whether there are control-flow graphs similar to the control-graph of the software, for which graph similarity computation is needed. More applications of graph distance can be found in~\cite{graphsim}.

Graph edit distance (GED) and maximum common subgraph (MCS) are two general measures for the similarity between two graphs~\cite{raymond2002rascal}. GED is the minimum number of edit operations (e.g., node/edge deletion/insertion) to transform one graph into another. MCS is the size of the largest common subgraph  (with respect to the number of nodes) shared by two graphs. Computing the exact GED and MCS between two graphs is known to be NP-hard and still challenging in practice~\cite{bunke1998graph, zeng2009comparing}. Moreover, it is reported that the state-of-the-art algorithms fail to compute the exact GED between 2 graphs with more than 16 nodes in a reasonable time~\cite{blumenthal2018exact}.

Many methods have been proposed to compute graph similarity, and they usually provide approximate results for computation speedup. These methods can be roughly classified into two categories, i.e., graph theory based methods and learning based methods. In the graph theory based methods, BEAM~\cite{neuhaus2006fast} uses beam search to avoid the high complexity for searching the full space. Hungarian~\cite{riesen2009approximate} and VJ~\cite{fankhauser2011speeding} use linear programming to approximate GED. HED~\cite{fischer2015approximation} matches the nodes in two graphs using their local structures. MC-SPLIT~\cite{a2017partition} uses a branch and bound algorithm to compute MCS. In the learning based methods, GraphSIM~\cite{graphsim} utilizes the graph convolutional network (GCN)~\cite{GCN} to compute the node embeddings and the embedding correlation matrix used to predict graph similarity. Graph matching network (GMN)~\cite{li2019graph} adopts an attention layer to match the nodes in two graphs in embedding learning and computes the GED using the embedding of the two graphs. Currently, the learning based methods are shown to outperform the graph theory based methods in both accuracy and efficiency, and thus benefit tasks that require graph similarity estimation. We will give a more detailed introduction to the related work and discuss the differences of HGMN from them in Section~\ref{sec:background}.     

%the A* algorithm builds a search tree to find the optimal solution~\citep{hart1968formal}, in which the root node represents one of the two graphs, the internal nodes correspond to partial solutions (after some editions) and the leaf nodes represent the other graph. As A* has exponential time complexity, many other algorithms are proposed, including BEAM~\cite{neuhaus2006fast},  Hungarian~\cite{riesen2009approximate}, VJ~\cite{fankhauser2011speeding} and HED~\cite{fischer2015approximation}. Besides, MC-SPLIT~\cite{a2017partition} is a branch and bound algorithm which is suitable for measuring MCS. It discovers a maximum-cardinality mapping from graphs to induce the result. However, traditional graph theory based methods are generally slower and only generate the approximate solution, compared with the methods with deep learning approaches. In learning based methods, GraphSIM\cite{bai2020learning} utilizes the graph convolutional network (GCN) to compute the node embeddings and predicts GED and MCS using the embedding correlation matrix of the two graphs as input. Graph matching network (GMN)\cite{li2019graph} adopts an attention layer to match the nodes in two graphs in embedding learning and computes the GED using the embedding of the two graphs. We will give a more detailed introduction to the related work in Section~\ref{sec:background}.

%   However, according to the topological structure of graphs, it will make local structural information lost in this situation, and the relation between their compact graphs will be hidden. 

Existing learning based methods either use the embedding of each individual node or the embedding of an entire graph~\cite{simgnn, li2019graph}, which fail to capture local topological structures of different scales. In this paper, we observe that graph similarity can benefit from a multi-scale view. That is, if two graphs are similar to each other, they are also similar when compressed into more compact graphs and conversely if two graphs are different their compact graphs are also likely to be different. We propose the hierarchical graph matching network (HGMN), which uses multiple stages of spectral clustering to cluster the graphs into successively more compact graphs. In each stage of the clustering, earth mover distance (EMD)~\cite{rubner1998EMD} is used to explicitly align the nodes in the two graphs such that the network does not have to learn complex node permutations. We derive correlation matrices from the node embedding in each stage and these matrices are fed into a convolutional neural network (CNN) to predict graph similarity. The entire pipeline is trained end-to-end in a data-driven fashion.

We experimented on 4 datasets (i.e., AIDS, LINUX, IMDB-MULTI,
and PTC) and used 4 performance matrices (i.e., mean
squared error, spearman’s rank correlation coefficient, kendall’s rank correlation coefficient, precision at $k$) to evaluate the accuracy of graph similarity approximation. The results show that HGMN consistently outperforms the state-of-the-art baselines on different datasets and performance metrics. Compared with the best performing baseline, the improvement in accuracy is 12.0\% on average and can be up to 62.6\%. Moreover, we also experimented the key designs in HGMN, i.e., hierarchical graph clustering and explicit node matching, and the results show that both of them lead to performance improvement. 

\section{Background and Related Work}\label{sec:background}
\begin{figure}[!t]	
	\centering
	\includegraphics[width=0.4\textwidth]{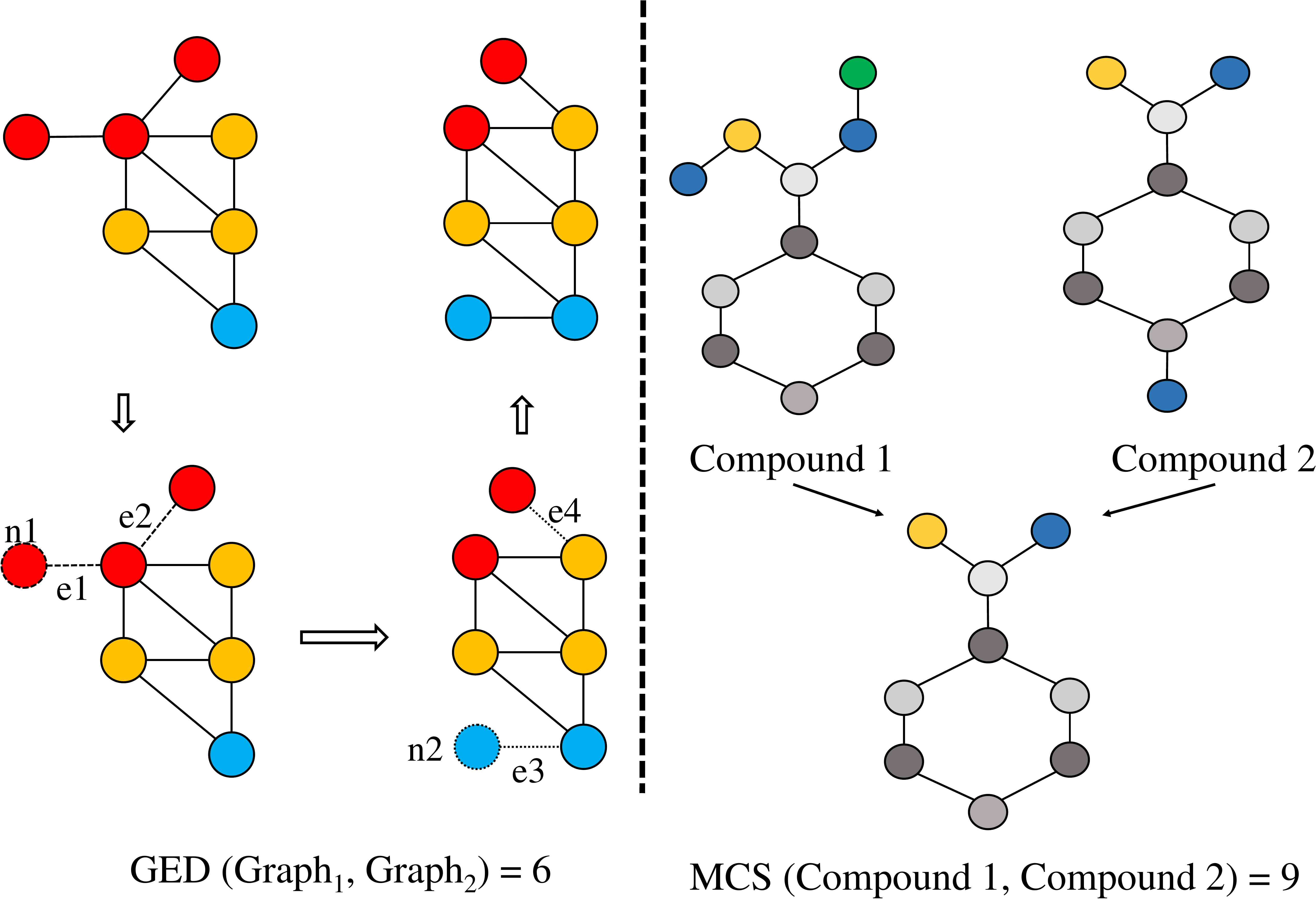}
	\caption{An illustration of GED and MCS, best viewed in color} 
	\label{fig:GED and MSC}
\end{figure}

\begin{figure*}[!t]	
	\centering
	\includegraphics[width=\textwidth]{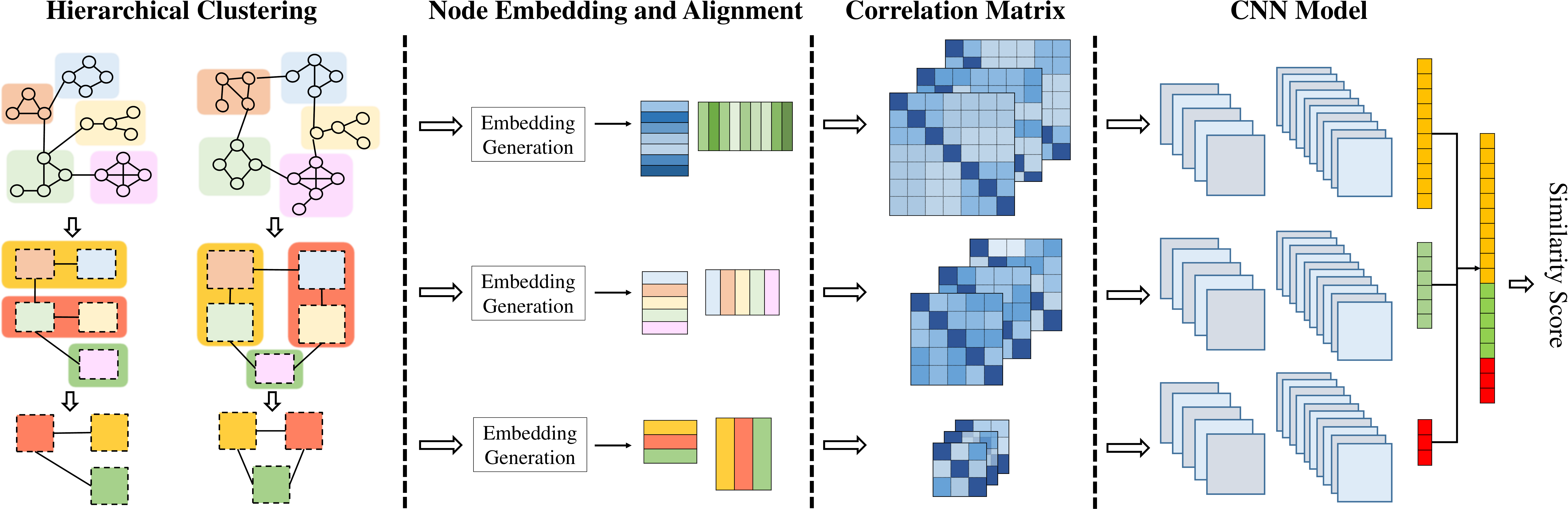}
	\caption{The pipeline of hierarchical graph matching network (HGMN), best viewed in color} 
	\label{fig:pipleline}
\end{figure*}

\subsection{Problem Formulation}

A graph is represented as $\mathcal{G}=\{\mathcal{V},\mathcal{E}\}$, in which $\mathcal{V}$ is the set of nodes and $\mathcal{E}$ is the set of edges. We denote $|\mathcal{V}|=N$ and $|\mathcal{E}|=M$. Each node $v\in \mathcal{V}$ can come with a feature vector $h_v$. We set $h_v$ as a one-hot vector or all 1 vector in $\mathbb{R}^N$ when the graph does not come with node feature vectors. We assume that edges do not have weight and focus on undirected graphs. The adjacency matrix of the graph is denoted using $\mathbf{A}$.

For two graphs $\mathcal{G}_1$ and $\mathcal{G}_2$, their GED is the minimum number of edit operations in the optimal alignment that transform one graph into another~\cite{zeng2009comparing}. The edit operations include edge deletion/insertion, node deletion/insertion and relabeling a node. We transform GED into a similarity score in the range of $(0, 1]$ using $e^{-GED(\mathcal{G}_1, \mathcal{G}_2)}$ to ensure that its range is well-defined. A maximum common subgraph of two graphs $\mathcal{G}_1$ and $\mathcal{G}_2$ is a subgraph common to both $\mathcal{G}_1$ and $\mathcal{G}_2$, and there is no other common subgraph of $\mathcal{G}_1$ and $\mathcal{G}_2$ that contains more nodes~\cite{raymond2002rascal}. We restrict the the maximum common subgraph to be a connected graph and MCS is defined as the number of nodes in the maximum common subgraph. An illustration of GED and MCS is provide in Figure~\ref{fig:GED and MSC}.

Our goal is to learn a model $f(\mathcal{G}_1, \mathcal{G}_2)$ that predicts the (transformed) GED or MCS between two graphs. Under reasonable computational complexity constraint, we want the prediction of the model to be as accurate as possible. 

%The prediction of the model is usually approximate but can significantly speed up graph distance (or similarity) computation.  

\subsection{Existing Methods}  

\noindent\textbf{Graph-theory based methods.} The A* algorithm~\cite{hart1968formal} is widely used for GED computation, which returns the exact result. A* is a best-first algorithm that tries to find an optimal path in a search tree. The idea is to formulate the problem using a tree structure, in which the root node is the starting point, inner nodes represent partial solutions, and leaf nodes are complete solutions. As A* has exponential time complexity, it is only suitable for small graphs and cannot finish in a reasonable time for large graphs. Several heuristics have been proposed to improve the execution time, which sacrifice accuracy for efficiency. For example, BEAM utilizes beam search to avoid searching the full search space and introduces a heuristic rule to favor long partial edit paths over shorter ones~\cite{neuhaus2006fast}.

Some methods measure GED via linear programing~\cite{bougleux2017graph}. Based on bipartite graph matching, Hungarian~\cite{riesen2009approximate} and VJ~\cite{fankhauser2011speeding} replace the cost of editing a node by the cost of editing the 1-star graph centered at this node. The cost of substituting a star graph by another one is further expressed as the solution of a square linear sum assignment problem. HED~\cite{fischer2015approximation} matches the nodes in two graphs using their local structures, and GED is approximated by the Hausdorff distance~\cite{huttenlocher1993comparing} between the nodes in the two graphs. To compute MCS, a branch and bound algorithm is used in MC-SPLIT~\cite{a2017partition}, which induces the result by finding a maximum-cardinality mapping between the graphs.

\vspace{1mm}

\noindent\textbf{Learning-based methods.} The learning based methods usually use graph neural networks (such as GCN) to learn embedding and predict graph similarity using the embedding. SMPNN predicts graph similarity using a summation of the similarities between the nodes in the two graphs~\cite{SMPNN}. GCNMEAN and GCNMAX~\cite{GCNMAX} use GCN~\cite{GCN} to learn graph embedding and train a fully connected neural network to compute graph similarity from the embedding of two graphs. SIMGNN uses both graph embedding and node similarities to predict graph similarity~\cite{simgnn}. GMN~\cite{li2019graph} introduces a cross-graph attention layer to allow the nodes in the two graphs to interact with each other but still predicts graph similarity using graph embedding. GraphSIM~\cite{graphsim} utilizes GCN with a different number of layers to build multiple correlation matrices among the nodes in the two graphs and use the correlation matrices to predict graph similarity. 

\vspace{1mm} 

\noindent\textbf{Our contributions.} HGMN adopts successful techniques from existing learning based methods, e.g., applying GCN to learn node embedding and using the node correlation matrix as input for neural network. However, HGMN has two fundamental differences from existing learning-based methods. First, we use multiple stages of spectral clustering to create a multi-scale view of the similarity between graphs. The hierarchical clustering provides more information for the downstream neural network as the differences between two graphs can be captured in the correlation matrices of different scales. Second, we explicitly align the nodes in the two graphs using the earth mover distance and computes correlation matrix in the aligned order. Node alignment ensures that the correlation matrix is the same under arbitrary node permutation and thus the neural network does not need to learn to be robust to node permutation. We will show in the experiments that both designs are crucial for performance, especially on large graphs.

%The definition of efficient and general similarity or dissimilarity measures between attributed graphs is a key problem in structural pattern recognition and graph based mining \cite{1979error, bunke1983inexact, sanfeliu1983distance, neuhaus2007bridging, neuhaus2007bridging, sanchez2012enabling}. As the ground truth we used in this work, this problem is nicely addressed by the graph edit distance (GED), which constitutes one of the most flexible dissimilarity measures between attributed graphs \cite{bougleux2017graph}. GED is generated from a sequence of edit operations to transform a graph to another. So the edit operations are restricted to be elementary: insertion or removal of a node or an edge. Computing the GED is thus a minimal cost problem. The cost means the minimal edit operations to achieve the graph transformation, which measures the dissimilarity between graphs.

\section{Hierarchical Graph Matching Network}

The data processing pipeline of HGMN is illustrated in Figure~\ref{fig:pipleline}. HGMN uses multiple stages of spectral clustering to organize the graphs into successively more compact graphs. In each stage, an embedding pooling operator is applied to derive the initial node embedding in this stage from the node embedding of the previous stage. Then, the initial node embeddings are processed by a GCN to generate refined node embedding. Based on the refined embedding, we use the earth mover distance to build a one-to-one mapping for nodes in the two graphs (i.e., node alignment) to ensure permutation invariance. The correlation matrices from all stages are fed into a CNN model to predict the similarity score of two graphs. The GCNs for all hierarchical clustering stages and the CNN model is trained from data. In the following, we introduce the modules of the HGMN pipeline in more details.

\subsection{Hierarchical Graph Clustering}

The procedure of hierarchical graph clustering is described in Algorithm~\ref{alg:graph compaction}, which is conducted in $L$ stages. In each stage, a new graph $\mathcal{G}^l$ and its adjacency matrix $\mathbf{A}^l$ are constructed from the graph in the previous stage (i.e., $\mathcal{G}^{l-1}$). The size sequence of the graphs satisfies $N>s_1>s_2>\cdots>s_L$, in which $N$ is the size of the original graph, $s_l$ is the size of the compact graph in $l\textsubscript{th}$ stage and we always use $s_L=1$. Therefore, the graph becomes smaller and smaller as the stage goes on. In the 4\textsuperscript{th} line of Algorithm~\ref{alg:graph compaction}, the normalized Laplacian $\tilde{\mathbf{L}}$ for a graph $\mathcal{G}$ with adjacency matrix $\mathbf{A}$ is defined as $\tilde{\mathbf{L}}=\mathbf{D}^{-\frac{1}{2}} \mathbf{L} \mathbf{D}^{-\frac{1}{2}}$, in which $\mathbf{D}$ is the diagonal degree matrix with $\mathbf{D}[i][i]=\sum_{j=1}^{N}\mathbf{A}[i][j]$ and $\mathbf{L}=\mathbf{D}-\mathbf{A}$ is the Laplacian matrix. Each row (i.e., $f^{l-1}_v$) of the matrix $F^{s_{l-1}\times k}$ corresponds to a node $v$ in $\mathcal{G}^{l-1}$. The k-means in the 7\textsuperscript{th} line of Algorithm~\ref{alg:graph compaction} groups the nodes in $\mathcal{G}^{l-1}$ into clusters and we treat each cluster of nodes as a single node in the compact graph $\mathcal{G}^l$. The for-loop in the 9\textsuperscript{th} line of Algorithm~\ref{alg:graph compaction} constructs the adjacency matrix of $\mathcal{G}^l$ and we assume two nodes are connected in $\mathcal{G}^l$ if they contain connected nodes $\mathcal{G}^{l-1}$.

\begin{algorithm}
	\caption{Hierarchical graph compaction with spectral clustering }
	\label{alg:graph compaction}
	\begin{algorithmic}[1]
		\STATE {\bfseries Input:} A graph $\mathcal{G}^0$, its adjacency matrix $\mathbf{A}^0$, the number of clustering stages $L$ and the size of the compact graphs for each stage $s_1,s_2,\cdots,s_L$
		\STATE {\bfseries Output:} $L$ successively more compact graphs $\mathcal{G}^1, \mathcal{G}^2,\cdots,\mathcal{G}^L$ and their adjacency matrices $\mathbf{A}^1,\mathbf{A}^2,\cdots,\mathbf{A}^L$ 
		\FOR{$1\le l\le L$}
		\STATE Compute the normalized Laplacian of $\mathcal{G}^{l-1}$ as $\tilde{\mathbf{L}}$ 
		\STATE Compute the eigenvectors corresponding to the $k$ smallest eigenvalues of $\tilde{\mathbf{L}}$ and use them as the columns of $F^{s_{l-1}\times k}$
		\STATE Normalize each row $f^{l-1}_v$ of matrix $F^{s_{l-1}\times k}$ to unit norm
		\STATE Conduct k-means clustering to cluster the rows of $F^{s_{l-1}\times k}$ into $s_l$ clusters $\mathcal{C}^l_1,\mathcal{C}^l_2,\cdots,\mathcal{C}^l_{s_l}$
		\STATE Initialize $\mathbf{A}^l=\{0\}^{s_l \times s_l}$
		\FOR{each edge $e_{uv}\in \mathcal{G}^{l-1}$}
		\IF{$f^{l-1}_u \in \mathcal{C}^l_{i}$, $f^{l-1}_v \in \mathcal{C}^l_{j}$ and $i\neq j$}
		\STATE $\mathbf{A}^l[i][j]=1$
		\ENDIF
		\ENDFOR 
		\ENDFOR	
	\end{algorithmic}
\end{algorithm}

We use spectral clustering for graph compaction for two reasons. Firstly, it preserves the local structure of the graph. As we focus on unweighted graphs, spectral clustering approximately minimizes the number of cross edges (normalized by the size of the graph clusters) between the graph clusters $\mathcal{C}^l_1,\mathcal{C}^l_2,\cdots,\mathcal{C}^l_{s_l}$. This means that nodes in each graph cluster tend to be strongly connected. Secondly, spectral clustering also allows flexible control of the number of graph clusters by setting the number of k-means centers. We provide an illustration of hierarchical graph clustering in the leftmost part of Figure~\ref{fig:pipleline}, which shows that well-connected nodes are grouped into the same graph cluster. Moreover, it can be observed that for two different graphs, their compact graphs in the same clustering stage are also different. Thus, hierarchical graph clustering provides the down stream neural network a multi-scale view of the differences between two graphs, which makes the task of graph similarity prediction easier. Hierarchical clustering also makes HGMN more expressive and general than existing learning-based methods for graph similarity approximation. As we use the original graph in the $0^{\text{th}}$ stage and set $s_L=1$ for the final stage, methods that use either node embedding or graph embedding can be regraded as special cases of HGMN.   

\vspace{1mm}

\textbf{Embedding pooling.} In each stage of hierarchical graph clustering, we derive the initial node embedding for $\mathcal{G}^{l}$ from the node embedding of $\mathcal{G}^{l-1}$. We call this procedure embedding pooling, which is motivated by EigenPooling~\cite{ma2019graph}. We show how embedding pooling works for a graph cluster $\mathcal{C}^l_k$, which contains $n$ nodes from $\mathcal{G}^{l-1}$ and is treated as a single node in $\mathcal{G}^{l}$. Assume that the node embedding of $\mathcal{G}^{l-1}$ has $d$-dimension, the embedding matrix for $\mathcal{C}^l_k$ can be organized as $\mathbf{H}^l_k\in \mathbb{R}^{|\mathcal{C}^l_k|\times d}$, in which each row corresponds to the embedding of a node from $\mathcal{C}^l_k$ in $\mathcal{G}^{l-1}$. We can also define an adjacency matrix $\mathbf{A}^l_k$ for the nodes in $\mathcal{C}^l_k$ by connecting edges in $\mathcal{G}^{l-1}$ for which both end points are contained in $\mathcal{C}^l_k$. With the adjacency matrix $\mathbf{A}^l_k$, we can define the Laplacian matrix $\mathbf{L}^l_k$ for $\mathcal{C}^l_k$ and solve the eigenvectors of the Laplacian matrix as $[u^l_k(1),u^l_k(2),\cdots, u^l_k(n)]$, in which $u^l_k(1)\in\mathbb{R}^{n}$ is the eigenvector corresponding to the largest eigenvalue of $\mathbf{L}^l_k$. The initial embedding vector of $\mathcal{C}^l_k$ in $\mathcal{G}^{l}$ is obtained as
\begin{equation}
h^l_k=u^l_k(1)^{\top} \mathbf{H}^l_k,
\end{equation} 
in which $h^l_k\in\mathbb{R}^{d}$ and $h^l_k$ is used as the initial embedding for node $\mathcal{C}^l_k$ in $\mathcal{G}^{l}$. The intuition is that $u^l_k(1)$ corresponds to high frequency signal on $\mathbf{A}^l_k$ in spectral graph theory. By projecting $\mathbf{H}^l_k$ onto $u^l_k(1)$, we keep the signal component in  $\mathbf{H}^l_k$ that changes the fastest on $\mathbf{A}^l_k$. Using more eigenvectors (e.g., $u^l_k(2)$, $u^l_k(3)$), we can create multiple initial embedding for $\mathcal{C}^l_k$ and these embedding can work in parallel in a similar manner to multiple image channels in a CNN. Figure~\ref{fig:pipleline} (third column) shows that multiple initial embedding can be used to generated multiple correlation matrices for a stage after they go through GCN update and node alignment.

\subsection{Node Embedding and Alignment}

We use a graph convolutional neural network (GCN)~\cite{GCN} to refine the initial embedding for each stage and the GCNs for different stages have the same number of layers but do not share the model parameters. Assume that the graph $\mathcal{G}$ contains  $N$ nodes, the adjacency matrix of $\mathcal{G}$ is $\mathbf{A}\in\mathbb{R}^{N\times N}$ and the initial embedding matrix is $\mathbf{H}\in\mathbb{R}^{N\times d}$. A layer of GCN updates the embedding as follows
\begin{equation}\label{equ:gcn}
\mathbf{H}'=\delta\big(\tilde{\mathbf{D}}^{-\frac{1}{2}}\tilde{\mathbf{A}} \tilde{\mathbf{D}}^{-\frac{1}{2}} \mathbf{H} \mathbf{W} \big),
\end{equation} 
in which $\delta(\cdot)$ is the activation function, $\tilde{\mathbf{A}}=\mathbf{A}+\mathbf{I}_N$ is the augmented adjacency matrix with self-loop, $\tilde{\mathbf{D}}$ is the degree matrix defined on the augmented adjacency matrix $\tilde{\mathbf{A}}$, $\mathbf{W}\in\mathbb{R}^{d\times d'}$ is a learnable mapping matrix, and $d'$ is the dimension of the embedding for $\mathbf{H}'$. The layer in equation~\ref{equ:gcn} can be stacked to form a multi-layer GCN. GCN is shown to achieve good performance on many graph-based tasks such as node classification~\cite{GCN}, link prediction~\cite{schlichtkrull2018modeling} and graph classification~\cite{schlichtkrull2018modeling}. Recently, it is also shown that GCN can approximate the Weisfeiler-Lehman (WL) graph isomorphism test~\cite{keyulu19how}, which decides whether two graphs are topologically
identical. We choose GCN as the default embedding model for HGMN due to its expressiveness and simplicity but more sophisticated graph neural network models such as GAT~\cite{GAT} and JK-Net~\cite{xu2018representation} can also be easily incorporated into HGMN.

\begin{figure}[!t]	
	\centering
	\includegraphics[width=0.5\textwidth]{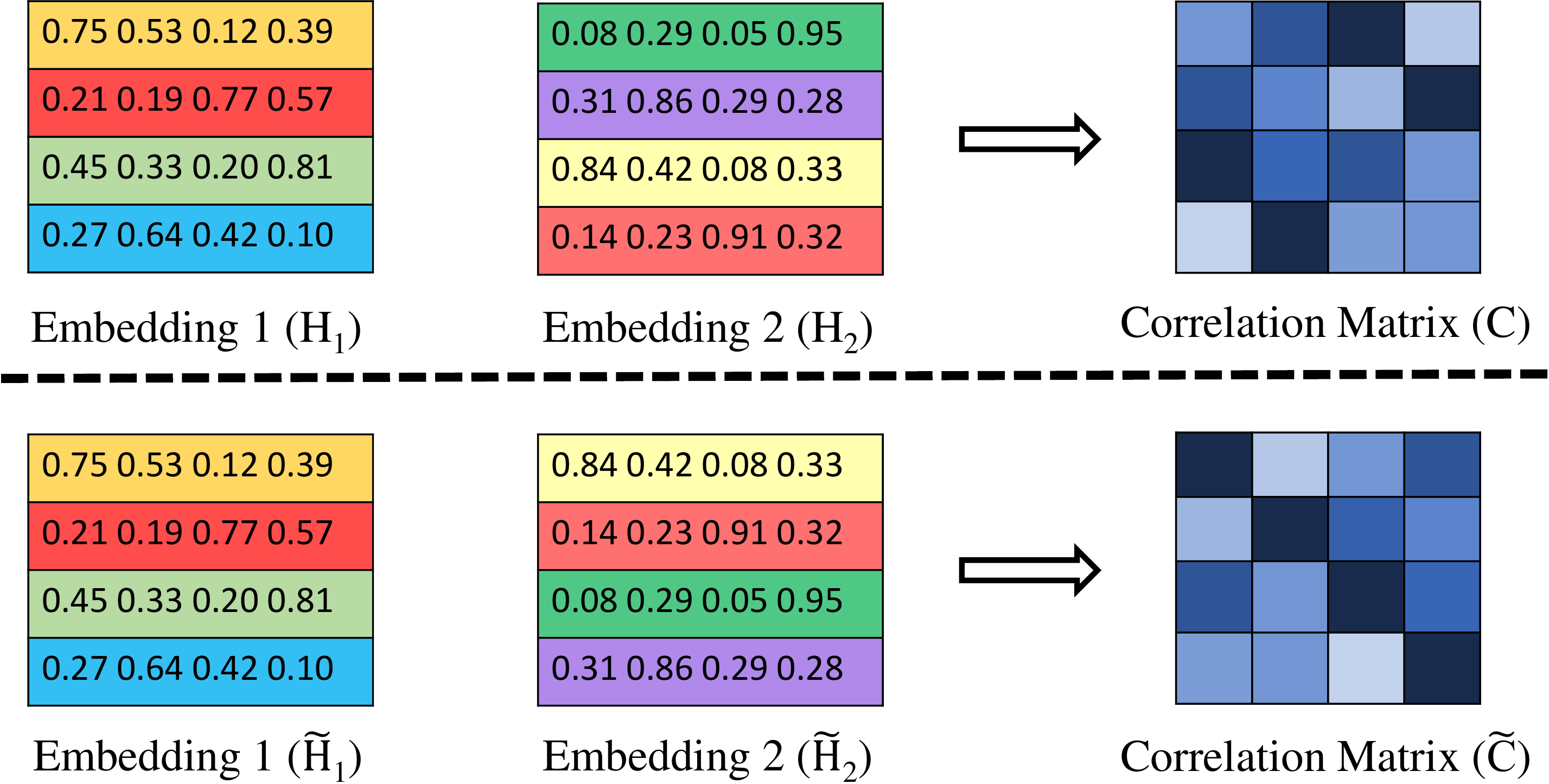}
	\caption{An illustration of the need for node alignment, different shades in $\mathbf{C}$ indicate different correlation values, best viewed in color} 
	\label{fig:order}
\end{figure}

% under a permutation of $\tilde{\mathbf{H}}_2$
Similar to GRAPHSIM~\cite{graphsim}, we use the embedding correlation matrices as the input for that neural network that predicts graph similarity. Assume that for a stage, we have two graphs $\mathcal{G}^1$ and $\mathcal{G}^2$ containing $N$ and $M$ nodes, respectively\footnote{Actually, only in the $0^{\text{th}}$ stage (i.e., on the original graphs), the two graphs can have different sizes, i.e., $N\neq M$. In each stage of clustering, the same output graph size is pre-specified for all graphs (i.e., $M=N$) such that the input matrices to the downstream CNN have fixed size. We use $M$ and $N$ for the size for $\mathcal{G}^1$ and $\mathcal{G}^2$ to consider the most general case.}. Their GCN embedding are denoted as $\mathbf{H}_1\in\mathbb{R}^{N\times d}$ and $\mathbf{H}_2\in\mathbb{R}^{M\times d}$ and the correlation matrix is $\mathbf{C}=\mathbf{H}_1\mathbf{H}_2^{\top}\in\mathbb{R}^{N\times M}$. However, as there is no canonical ordering of the nodes in a graph, the rows of $\mathbf{H}_1$ and $\mathbf{H}_2$ will be permuted under different node numbering, which results in different $\mathbf{C}$. We provide an illustration of this phenomenon in Figure~\ref{fig:order}, in which $\mathbf{H}_1=\tilde{\mathbf{H}}_1$ and $\tilde{\mathbf{H}}_2$ is a permutation of $\mathbf{H}_2$. We want $\mathbf{C}$ and $\tilde{\mathbf{C}}$ to be identical as $(\mathbf{H}_1, \mathbf{H}_2)$ and $(\tilde{\mathbf{H}}_1, \tilde{\mathbf{H}}_2)$ essentially represent the same pair of graphs. However, as shown in Figure~\ref{fig:order}, $\mathbf{C}$ and $\tilde{\mathbf{C}}$ are quite different. This means that the downstream CNN needs to be robust to node permutation, which makes the learning task difficult. Therefore, we use the earth mover distance~\cite{rubner1998EMD} to explicitly align the nodes in $\mathcal{G}^1$ and $\mathcal{G}^2$.

We define a distance matrix $\mathbf{D}\in \mathbb{R}^{N\times M}$ on $\mathbf{H}_1$ and $\mathbf{H}_2$ as $\mathbf{D}(i,j)=\Vert \mathbf{H}_1(i)- \mathbf{H}_2(j)\Vert_2$, in which $\mathbf{H}(k)$ denotes the $k^{\text{th}}$ row of matrix $\mathbf{H}$. Then the earth mover distance between $\mathbf{H}_1$ and $\mathbf{H}_2$ is defined as follows
\begin{align}\label{eq:3}
d(\mathbf{H}_1,\mathbf{H}_2) :=\min_{\mathbf{W}\in \mathcal{U}}  \sum_{i=1}^{N}\sum_{j=1}^{M}\mathbf{W}(i,j)\mathbf{D}(i,j),
\end{align}
in which $\mathcal{U}=\left \{ \mathbf{W}\in \mathbb{R}_+^{N\times M} \;|\;  \mathbf{W}\mathds{1}=\frac{1}{N}\mathds{1}, \mathbf{W}^T \mathds{1} =\frac{1}{M}\mathds{1} \right\}$. Intuitively, $\mathbf{D}(i,j)$ models the cost of transporting unit mass from $\mathbf{H}_1(i)$ to $\mathbf{H}_2(j)$ while $\mathbf{W}(i,j)$ models the amount of mass transported from $\mathbf{H}_1(i)$ to $\mathbf{H}_2(j)$. As $\mathbf{W}$ can be marginalized into $\frac{1}{N}\mathds{1}$ and $\frac{1}{M}\mathds{1}$, each row of $\mathbf{H}_1$ and $\mathbf{H}_2$ will send/receive $\frac{1}{N}$ and $\frac{1}{M}$ unit of the mass, respectively. By minimizing over $\mathbf{W}$, the earth mover distance encourages $\mathbf{W}(i,j)$ to be large if the distance between $\mathbf{H}_1(i)$ and $\mathbf{H}_2(j)$ is small.

Algorithm~\ref{alg:node matching} shows how to obtain a node matching between two graphs using the weight matrix $\mathbf{W}$ optimized by the earth mover distance. The idea is trying to match node pairs with large $\mathbf{W}(i,j)$ in a greedy fashion. In the 5\textsuperscript{th} line of Algorithm~\ref{alg:node matching}, ties are broken by taking the solution with the minimum value if there are multiple optimal solutions. As $\mathbf{W}(i,j)$ is likely to be large when $\mathbf{H}_1(i)$ and $\mathbf{H}_2(j)$ have a small distance, Algorithm~\ref{alg:node matching} essentially matches a node in $\mathcal{G}^1$ to a similar node in  $\mathcal{G}^2$. If we define an ordering for the nodes in $\mathcal{G}^1$, e.g., in descending order of the first dimension of their embedding, and arrange the nodes in $\mathcal{G}^2$ using the matched order, the correlation matrix will be the same as long as $(\mathbf{H}_1, \mathbf{H}_2)$ can be transformed into $(\tilde{\mathbf{H}}_1, \tilde{\mathbf{H}}_2)$ via permutation. Therefore, with explicit node alignment, the downstream CNN does not need to be robust to node permutation, which simplifies learning.

\begin{algorithm}
	\caption{Node matching with earth mover distance}
	\label{alg:node matching}
	\begin{algorithmic}[1]
		\STATE {\bfseries Input:} The size $M$ and $N$ of the two graphs $\mathcal{G}^1$ and $\mathcal{G}^2$, and the weight matrix $\mathbf{W}$ 
		\STATE {\bfseries Output:} A matching vector $T$, in which $T(i)$ is the match of node $i\in \mathcal{G}^1$ in graph $\mathcal{G}^2$ 
		\STATE Initialize $\mathcal{S}=\{1,2,\cdots, M\}$  
		\FOR{$1\le i \le \min (M,N)$}
		\STATE $j=\arg \max_{k\in\mathcal{S}} \mathbf{W}(i,k)$
		\STATE $T(i)=j$
		\STATE Delete $j$ from $\mathcal{S}$
		\ENDFOR	
	\end{algorithmic}
\end{algorithm}

One subtlety is that the correlation matrix $\mathbf{C}\in\mathbb{R}^{N\times M}$ does not have a fixed size for the original graphs (the $0^{\text{th}}$ stage of clustering). To ensure that the CNN has fixed size input, we use interpolation to up-sample $\mathbf{C}$ to a size of $P\times P$ for the $0^{\text{th}}$ stage, in which $P$ is the size of the largest graph in the dataset.

\subsection{Network Structure and Loss Function} 

Our network uses the correlation matrices from all stages as input to predict graph similarity. As illustrated in the rightmost part of Figure~\ref{fig:pipleline}, the network consists of multiple convolutional layers and fully connected layers. Since the correlation matrices are similar to images, the convolutional layers are utilized to extract spatial features from them. The fully connected layers allow the features from different stages to interact with each other. The output of the network is a single value that indicates the distance / similarity between two graphs.

We formulate graph similarity prediction as a regression problem and use the mean squared error as the loss function
\begin{equation}
L(w)=\frac{1}{B}\sum_{t=1}^{B} \big[f(\mathcal{G}^1_t, \mathcal{G}^2_t;w)-d(\mathcal{G}^1_t, \mathcal{G}^2_t)\big]^2,
\end{equation}    
in which $w$ is the model parameter, $f(\mathcal{G}^1_t, \mathcal{G}^2_t;w)$ is the graph distance predicted by the model for a graph pair and $d(\mathcal{G}^1, \mathcal{G}^2)$ is the ground-truth distance between the graph pair. We use min-batch stochastic gradient descent (SGD) for training and in each min-batch, $B$ graph pairs are randomly sampled from the training set to compute loss. The trainable parameters in the model include the CNN used for distance prediction and the GCNs used for embedding in all stages.

HGMN can compute graph similarity efficiently, especially for graph similarity search, in which the dataset is known before hand. Hierarchical graph clustering and node embedding can be conducted for the graphs in the database before the query comes.  Spectral clustering for the query graph has a complexity of $\mathcal{O}(P^3)$ and earth mover distance based node alignment has a complexity of $\mathcal{O}(P^2)$ if the largest query graph has a size of $P$. The $\mathcal{O}(P^3)$ complexity will not be a big problem if the graphs are not too large. For other computations that involve neural networks, the complexity of HGMN is similar to existing learning based methods such as GRAPHSIM~\cite{graphsim} and GMN~\cite{li2019graph}. 

\begin{table*}
	\centering
	\caption{Dataset statistics}
	\label{tab:dataset statistics}
	\begin{tabular}{l|ccccc}
		\toprule
		\thead{Dataset} & \thead{Domains} & \thead{\# Graphs} & \thead{MIN/MAX nodes per graph}  & \thead{AVG nodes per graph}  \\
		\midrule
		\thead{AIDS} & Chemical compounds & 700 & 2/10 & 8.9 \\
		\thead{LINUX} & Program dependence graph & 1000  & 4/10 & 7.6 \\
		\thead{IMDB-MULTI} & Ego-networks & 332  & 16/89 & 25.0 \\
		\thead{PTC} & Biochemistry & 256 & 16/103 & 30.2 \\
		\bottomrule
	\end{tabular}
\end{table*}             

%\begin{figure}[!t]	
%	\centering
%	\begin{subfigure}{0.48\linewidth}
%		\centering 
%		\includegraphics[width=\textwidth]{./figures/holder.jpg} 
%		\caption{Edge deletion}
%		\label{fig:delete edge}
%	\end{subfigure}
%	\begin{subfigure}{0.48\linewidth}
%		\centering 
%		\includegraphics[width=\textwidth]{./figures/holder.jpg} 
%		\caption{Edge addition}
%		\label{fig:add edge}
%	\end{subfigure}
%	\caption{Relation between noise ratio and classification accuracy for GCN, GAT and SGC on the xxx dataset, x-noise ratio, y-classification accuracy} 
%	\label{fig:noise ratio}
%\end{figure}

\section{Experimental Evaluation}\label{sec:experiment}

In this part, we first introduce the experiment settings, including datasets, performance metrics and baselines. Then we compare HGMN with the baselines for accuracy of graph similarity prediction. Finally, we examine the key designs in HGMN, i.e., hierarchical graph clustering and earth mover distance based node alignment, and test the influence of the parameters on the performance. All codes to reproduces the results of HGMN will be released after the review process.   

\subsection{Experimental Settings}

We largely follow the experimental settings in~\cite{graphsim} and introduce the details as follows.

\textbf{Datasets}. We used 4 real datasets for the experiments, i.e., \textit{AIDS, PTC, LINUX and IMDB-MULTI}. The \textit{AIDS} dataset contains 42,687 chemical compound graphs from the Developmental Therapeutics Program at NCI/NIH 7 and each node in a graph is associated with one out of 29 labels. AIDS has been widely used for the evaluation of graph similarity computation\cite{zeng2009comparing, zhao2013efficient, simgnn} and we randomly sampled 700 graphs from the dataset. \textit{PTC} consists of 344 chemical compound graphs that report the carcinogenicity for male and female rats. Each node in the PTC dataset has one out of 19 possible labels. \textit{LINUX} has 48,747 Program Dependence Graphs (PDG) generated from the Linux kernel. In each PDG, a node is one statement and an edge models the dependency between the two statements. We randomly sampled 1,000 graphs from the original LINUX dataset. \textit{IMDB-MULTI} is a movie-collaboration dataset containing 1,500 ego-networks of movie actors/actresses. In the ego-networks, each node represents a person and an edge models the collaboration between two persons. On IMDB-MULTI and PTC, we removed graphs containing less than 16 nodes to test the scalability of our methods. For both LINUX and IMDB-MULTI, the nodes do not come with a label and we use the all 1 vector as the initial embedding of the nodes for the two datasets. The statistics of the datasets after preprocessing are reported in Table~\ref{tab:dataset statistics}.

\vspace{1mm}

\textbf{Evaluation Methodology}. For each dataset, we generated the training set, validation set and test set with a split ratio of 7:2:1. The model was trained on the training set and the hyper-parameters (e.g., the number of stages in hierarchical graph clustering and the number of layers in GCN ) were tuned using the validation set. Graphs in the test set were treated as queries and we evaluated how accurately the model approximates the similarity between the query graphs and the graphs in the entire dataset. For AIDS and LINUX, the A* algorithm was used to compute the ground-truth GED between the graphs. As A* has an exponential time complexity with respect to the number of nodes in the graphs, it took too much time for PTC and IMDB-MULTI. Therefore, we computed the ground-truth GED for PTC and IMDB-MULTI by taking the minimum of three approximate algorithms, i.e., Beam~\cite{neuhaus2006fast}, Hungarian~\cite{riesen2009approximate} and VJ~\cite{fankhauser2011speeding}. The distances returned by the algorithms are larger than or equal to the true GED and the same ground-truth approximation methodology was also adopted in~\cite{graphsim}. To compute the ground-truth MCS, we used MC-SPLIT~\cite{a2017partition} as it can finish in a long but tolerable time for our datasets. Note that the algorithms used to provide the ground-truth distance/similarity are typically orders of magnitude slower than learning based methods~\cite{graphsim}. We excluded the test set from model training to show that the trained models can generalize to unseen data and thus improve the efficiency of graph similarity search.   

We used four performance metrics to evaluate the accuracy of graph similarity approximation, i.e., \textit{average mean squared error} (MSE), \textit{Spearman’s rank correlation coefficient} ($\rho$), \textit{Kendall’s rank correlation coefficient} ($\tau$) and \textit{precision at 10} ($p@10$). MSE is the mean squared error of the predicted GED/MCS compared with the ground-truth GED/MCS. For a query graph $\mathcal{G}$, the graphs in the dataset were ranked according to their predicted graph similarities. Both $\rho$ and $\tau$ evaluate how well the similarity prediction based ranking matches ground truth similarity based ranking, and higher value means better performance. $p@10$ is the percentage of true top-10 nearest neighbor in the top-10 nearest neighbors obtained from estimated graph similarity. MSE measures the accuracy of graph similarity approximation, while $\rho$, $\tau$ and $p@10$ evaluate how well the estimated graph similarity ranks the graphs, which is also important for graph similarity search.

\vspace{1mm}
\textbf{Baselines}. As the learning based methods were shown to outperform the graph theory based methods in both accuracy and efficiency~\cite{graphsim}, we mainly compared with the learning based methods. The baselines include SMPNN~\cite{SMPNN}, GCNMEAN, GCNMAX~\cite{GCNMAX}, SIMGNN~\cite{simgnn}, GMN~\cite{li2019graph} and GraphSIM~\cite{graphsim}. EMBAVG is a simple baseline introduced in~\cite{graphsim} that computes graph similarity using the dot product of two graph embeddings. As the results of our run for SIMGNN and GraphSIM on the AIDS and LINUX dataset are slightly worse than those reported in their papers, we reused the results from their papers. By default, HGMN uses 4 hierarchical cluster stages with size 6, 4, 2, 1 for AIDS and LINUX (small graphs), and 6 hierarchical cluster stages with size 64, 16, 8, 4, 2, 1 for PTC and IMDB-MULT (large graphs).

\begin{table*}[htbp]
	\caption {Accuracy comparison for GED approximation, for $mse$, smaller value means better performance, for $\rho$, $\tau$ and $p@10$, larger value means better performance, the last column if the improvement of HGMN over the best performing baseline}
	\label{tab:GED}
	
	\begin{tabular}{ccccccccc|c|c}
		\toprule
		
		\multicolumn{2}{c|}{Dataset and Metric} & \thead{SMPNN} & \thead{EMBAVG} & \thead{GCNMEAN} & \thead{GCNMAX} & \thead{SIMGNN} & \thead{GMN} & \thead{GRAPHSIM} & \thead{HGMN} & \thead{Gain (\%)} \\
		
		\midrule
		\midrule
		
		\multirow{4}{*}{AIDS} & \multicolumn{1}{r|}{$mse$} &4.725 & 3.185 & 2.124 & 3.423 & 1.189 & 1.741 & 0.787 & 0.752 & 4.4\\
		& \multicolumn{1}{r|}{$\rho$} & 0.306 & 0.642 & 0.653 & 0.628 & 0.843 & 0.751 & 0.874 & 0.883 & 1.0\\
		& \multicolumn{1}{r|}{$\tau $} & 0.480 & 0.592 & 0.629 & 0.505 & 0.690 & 0.642 & 0.776 & 0.778 & 0.3\\
		& \multicolumn{1}{r|}{$p @ 10 $} & 0.092 & 0.179 & 0.194 & 0.290 & 0.421 & 0.401 & 0.534 & 0.537 & 0.5\\
		
		\midrule
		
		\multirow{4}{*}{LINUX} & \multicolumn{1}{r|}{$mse$} & 11.523 & 11.244 & 7.541 & 6.341& 1.509& 1.027& 0.058 & 0.056 & 3.4\\
		& \multicolumn{1}{r|}{$\rho $}& 0.046 & 0.245 & 0.579 & 0.724 & 0.939 & 0.941 & 0.981 & 0.984 & 0.3\\
		& \multicolumn{1}{r|}{$\tau $} & 0.016 & 0.301 & 0.525 & 0.740 & 0.879 & 0.896 & 0.907 &0.920 & 1.4\\
		& \multicolumn{1}{r|}{$p @ 10 $} & 0.014 & 0.071 & 0.141 & 0.541 & 0.942 & 0.933 & 0.992 & 0.996 & 0.4\\
		
		\midrule
		
		\multirow{4}{*}{IMDB-MULTI} & \multicolumn{1}{r|}{$mse$} & 32.596 & 71.789 & 68.823 & 58.425 & 2.964 & 3.210 & 1.924 & 0.719 & 62.6\\
		& \multicolumn{1}{r|}{$\rho$} & 0.107 & 0.229 & 0.402 & 0.449 & 0.781 & 0.725 & 0.825 & 0.930 & 12.7\\
		& \multicolumn{1}{r|}{$\tau$} & 0.644 & 0.187 & 0.378 & 0.354 & 0.770 & 0.782 & 0.821 & 0.914 & 11.3\\
		& \multicolumn{1}{r|}{$p @ 10 $} & 0.021 & 0.210 & 0.219 & 0.437 & 0.724 & 0.751 & 0.813 & 0.853 & 4.9\\
		
		\midrule
		
		\multirow{4}{*}{PTC} & \multicolumn{1}{r|}{$mse $} &134.124 & 44.184 & 7.428 & 8.329 & 1.473 & 1.854 & 0.889 & 0.820 & 7.8\\
		& \multicolumn{1}{r|}{$\rho $} &0.127 & 0.324 & 0.546 & 0.506 & 0.726 & 0.670 & 0.714 & 0.958 & 34.2\\
		& \multicolumn{1}{r|}{$\tau $} & 0.167 & 0.315 & 0.490 & 0.468 & 0.678 & 0.592 & 0.719& 0.941 & 30.9\\
		& \multicolumn{1}{r|}{$p @ 10 $} & 0.087 & 0.144 & 0.210 & 0.241 & 0.475 & 0.374 & 0.541 & 0.623 & 15.2\\
		
		\bottomrule
	\end{tabular}
\end{table*}

\begin{table*}[htbp]
	\caption {Accuracy comparison for MCS approximation, for $mse$, smaller value means better performance, for $\rho$, $\tau$ and $p@10$, larger value means better performance, the last column if the improvement of HGMN over the best performing baseline}
	\label{tab:MCS}
	\begin{tabular}{ccccccccc|c|c}
		\toprule
		
		\multicolumn{2}{c|}{Dataset and Metric} & \thead{SMPNN} & \thead{EMBAVG} & \thead{GCNMEAN} & \thead{GCNMAX} & \thead{SIMGNN} & \thead{GMN} & \thead{GRAPHSIM} & \thead{HGMN} & \thead{Gain (\%)}\\

		\midrule
		\midrule
		
		\multirow{4}{*}{AIDS} & \multicolumn{1}{r|}{$mse $} & 4.268 & 6.148 & 6.234 & 4.156 & 3.433 & 2.234 & 2.402 & 2.213 & 0.9\\
		& \multicolumn{1}{r|}{$\rho $} & 0.772 & 0.723 & 0.756 & 0.801 & 0.822 & 0.901 & 0.858 & 0.902 & 0.1\\
		& \multicolumn{1}{r|}{$\tau $} & 0.529 & 0.510 & 0.498 & 0.574 & 0.680 & 0.803 & 0.798 & 0.871 & 9.1\\
		& \multicolumn{1}{r|}{$p @ 10$} & 0.379 & 0.243 & 0.347 & 0.315 & 0.374 & 0.513 & 0.505 & 0.525 & 2.3\\
		
		\midrule
		
		\multirow{4}{*}{LINUX} & \multicolumn{1}{r|}{$mse $} & 3.397 & 2.784 & 2.689 & 2.170 & 0.729 & 0.794 & 0.164 & 0.153 & 6.7\\
		& \multicolumn{1}{r|}{$\rho$} & 0.134 & 0.475 & 0.521 & 0.714 & 0.859 & 0.939 & 0.962 & 0.960 & 0.2\\
		& \multicolumn{1}{r|}{$\tau$} & 0.675 & 0.715 & 0.747 & 0.784 & 0.889 & 0.934 & 0.946 & 0.962 & 1.7\\
		& \multicolumn{1}{r|}{$p @ 10$} &  0.235 & 0.378 & 0.421 & 0.459 & 0.850 & 0.949 & 0.951 & 0.960 & 0.9\\
		
		\midrule
		
		\multirow{4}{*}{IMDB-MULTI} & \multicolumn{1}{r|}{$mse$} & 15.145 & 19.354 & 10.457 & 10.124 & 2.451 & 0.590 & 1.287 & 0.529 & 10.3\\
		& \multicolumn{1}{r|}{$\rho $} & 0.310 & 0.478 & 0.746 & 0.841 & 0.930 & 0.941 & 0.976 & 0.981 & 0.5\\
		& \multicolumn{1}{r|}{$\tau $} & 0.530 & 0.386 & 0.611 & 0.619 & 0.879 & 0.920 & 0.946 & 0.981 & 3.7\\
		& \multicolumn{1}{r|}{$p @ 10$} & 0.01 & 0.211 & 0.387 & 0.451 & 0.812 & 0.875 & 0.882 & 0.896 & 1.6\\
		
		\midrule
		
		\multirow{4}{*}{PTC} & \multicolumn{1}{r|}{$mse$} & 14.875 & 26.412 & 12.441 & 13.845 & 5.419 & 3.142 & 3.975 & 2.551 & 18.8\\
		& \multicolumn{1}{r|}{$\rho $} & 0.578 & 0.647 & 0.578 & 0.6617 & 0.712 & 0.782 & 0.779 & 0.811 & 3.7\\
		& \multicolumn{1}{r|}{$\tau $} & 0.522 & 0.419 & 0.650 & 0.688 & 0.746 & 0.792 & 0.8& 0.812 & 1.5\\
		& \multicolumn{1}{r|}{$p @ 10 $} & 0.187 & 0.352 & 0.384 & 0.402 & 0.356 & 0.584 & 0.498 & 0.609 & 4.3\\
		
		\bottomrule
	\end{tabular}
\end{table*}

\subsection{Main Performance Results}

We report our main results in Table~\ref{tab:GED} and Table~\ref{tab:MCS}, which compare HGMN with the baselines for the accuracy of GED and MCS perdition, respectively. We can make two observations from the results. First, HGMN consistently outperforms the baselines for both GED and MCS, across 4 different datasets and 4 different performance metrics. For GED, the performance improvement over the best performing baseline is 11.96\% on average (averaged over all datasets and performance metrics) and can be up to 62.6\%. Compared with the best performing baseline, the improvement for MCS is 4.15\% on average and can be up to 18.8\%. Second, the performance improvement of HGMN is significantly better for the larger datasets (IMDB-MULTI and PTC) than the smaller datasets (AIDS and LINUX). We conjecture that this is because larger graphs have richer structures when they are clustered into more compact graphs. These structures are better captured on the compact graphs than on the original graphs. In contrast, the graphs in AIDS and LINUX are small (with no more than 10 nodes) and thus GCN with a moderate number of layers is already able to capture the structures of different scales. For large graphs, GCN with a large number of layers are required but graph neural networks with too many layers are known to be prone to over-smoothing~\cite{chen2019measuring}, which often leads to poor performance. This explanation suggests that the hierarchical clustering may enable HGMN to perform well on even larger graphs and we provide more evidences to support this explanation in Section~\ref{subsec:ablation}.  

%We compared HGMN with the baselines for the accuracy of GED approximation in Table~\ref{tab:GED}. The results show that HGMN consistently improves the baselines on the 4 datasets and 4 performance metrics. Compared with the best performing baseline, the improvement is 12.0\% on average and can be up to 62.6\%. The performance results for MCS are reported in Table~\ref{tab:MCS} and similar to the case of GED, HGMN consistently outperforms the baselines on different datasets and performance metrics. The improvement over the best performing baseline is 4.2\% on average and can be up to 18.8\%. These results suggest that HGMN provides more accurate graph distance/similarity approximation than existing methods. We believe the good performance of HGMN mainly comes from its key designs, i.e., hierarchical graph clustering and explicit node matching, which are also the main differences between HGMN and existing methods.          

\begin{table*}[htbp]
	\centering
	\caption{Ablation study of the designs of HGMN for GED approximation}
	\label{tab:ablation}
	\begin{tabular}{l|cccc|cccc}
		\toprule
		
		\multirow{3}{*}{HGMN Variants}&
		\multicolumn{8}{c}{Performance Metric} \\
		& \multicolumn{4}{c|}{AIDS} 
		& \multicolumn{4}{c}{PTC} \\
		& \thead{$mse$} & \thead{$\rho$} & \thead{$\tau$} & \thead{$p@k$} & \thead{$mse$} & \thead{$\rho$} & \thead{$\tau$} & \thead{$p@k$} \\
		
		\midrule
		\midrule
		
		HGMN & 0.752 & 0.883 & 0.778 & 0.537 & 0.820 & 0.958 & 0.941 & 0.623 \\
		Without node alignment & 0.896 & 0.776 & 0.685 & 0.432 & 0.837 & 0.914 & 0.847 & 0.602\\
		Without hierarchical clustering & 2.768 & 0.612 & 0.605& 0.313 & 7.285 & 0.573 & 0.513 & 0.238 \\
		
		\bottomrule
	\end{tabular}
\end{table*}

\subsection{Ablation Study and Parameter Analysis}\label{subsec:ablation}

In Table~\ref{tab:ablation}, we study how the two key designs of HGMN, i.e., hierarchical clustering and node alignment, may influence the performance. For~\textit{without node alignment}, we used a random ordering of the nodes in the graphs. For~\textit{without hierarchical clustering}, we used only the embedding correlation matrix for the original graph, which is similar to the case of GRAPHSIM~\cite{graphsim}. We tested the influence of a dataset with small graphs (AIDS) and dataset with large graphs (PTC). The results show that disabling either node alignment or hierarchical clustering degrades the performance. Comparatively, the performance degradation is more severe for the large PTC dataset than the small AIDS dataset. This is another evidence that hierarchical clustering helps achieve good performance for large datasets. Compared with node alignment,  hierarchical clustering seems to be more important for the performance, and without it the $mse$ increases 2.7x and 7.9x for AIDS and PTC, respectively.

\begin{figure}[!t]	
	\centering
	\begin{subfigure}{0.6\linewidth}
		\centering 
		\includegraphics[width=\textwidth]{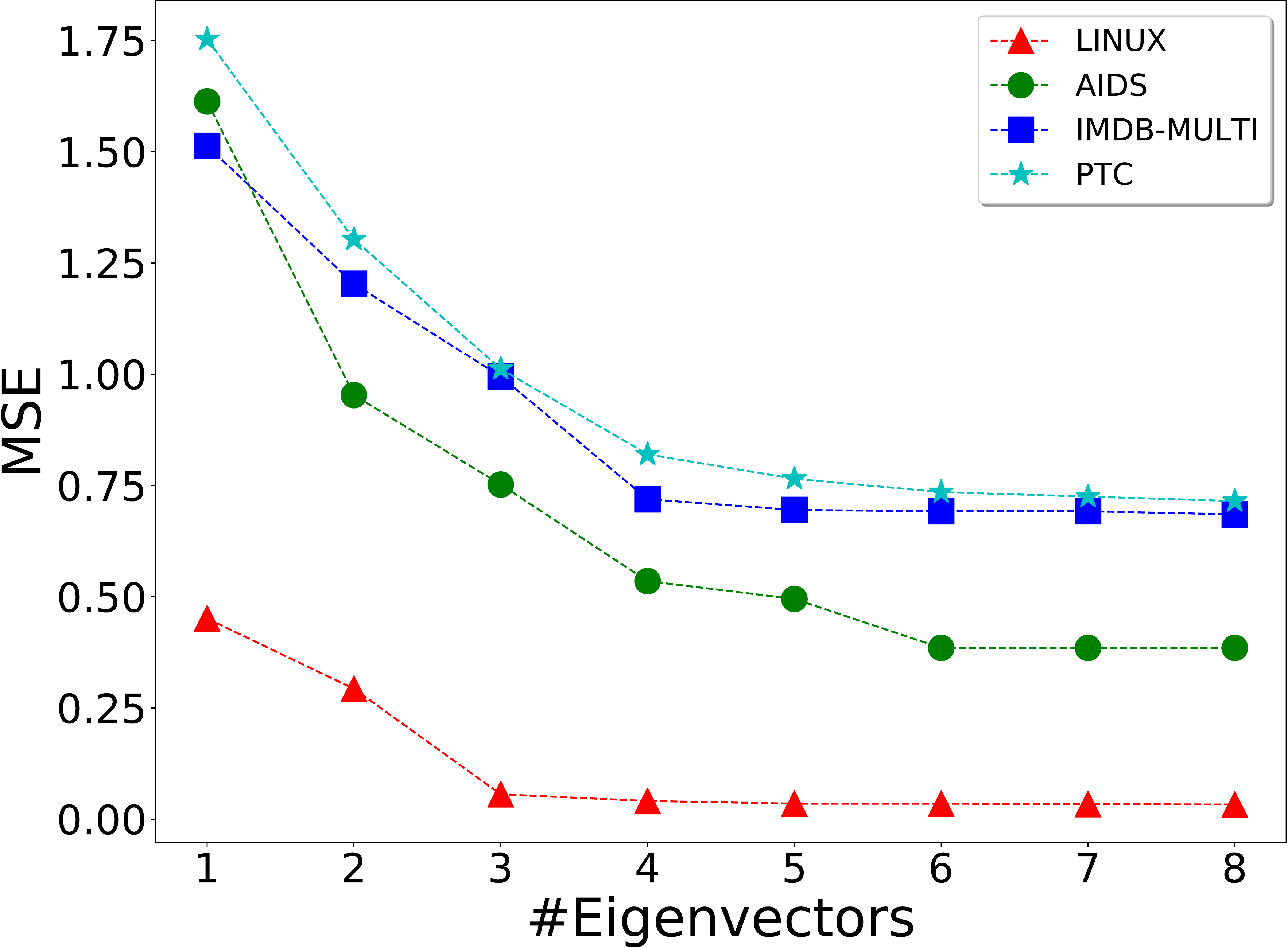} 
		\caption{\# eigenvector in pooling}
		\label{fig:vectors}
	\end{subfigure}\\
	\begin{subfigure}{0.6\linewidth}
		\centering 
		\includegraphics[width=\textwidth]{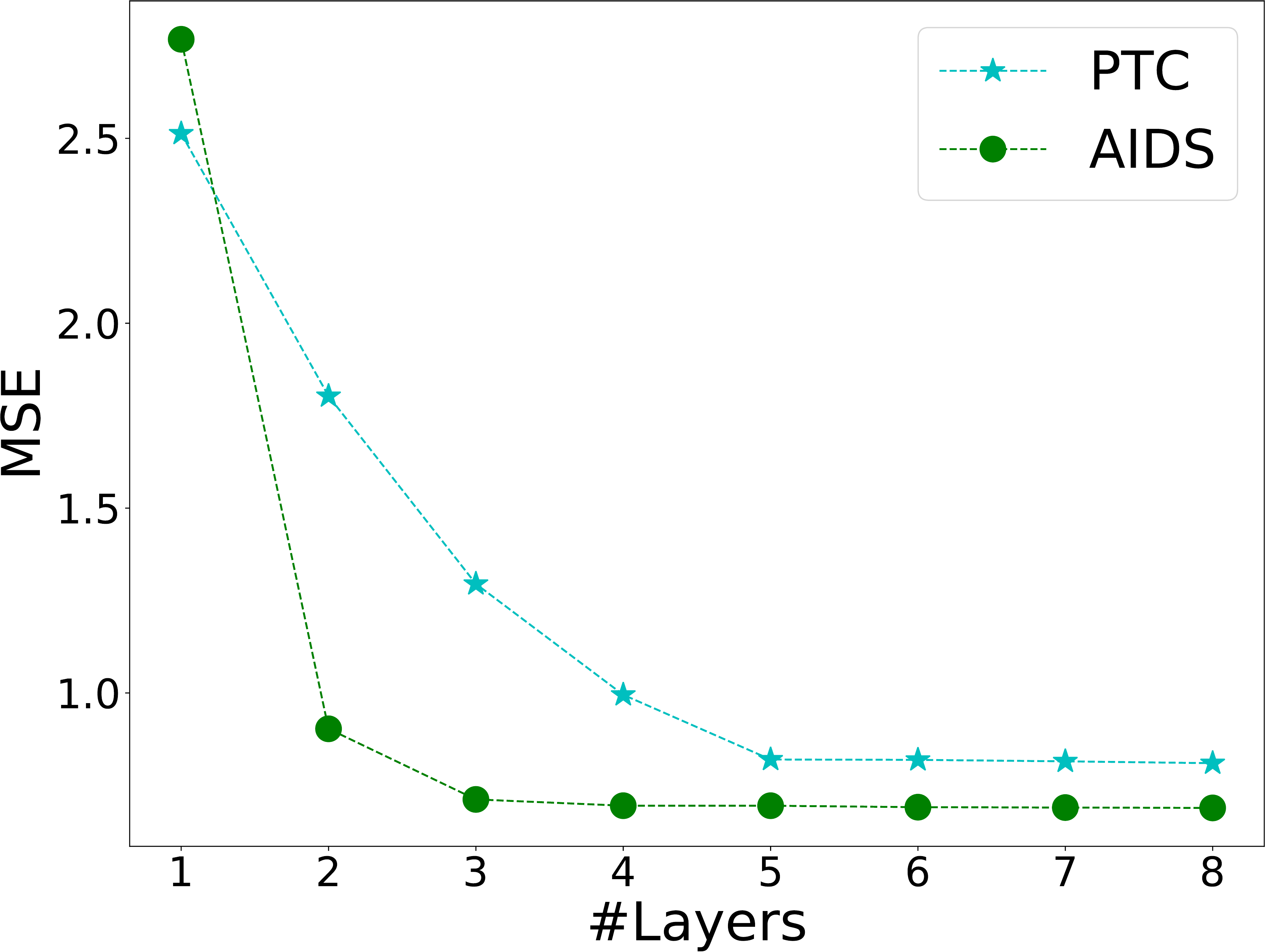} 
		\caption{\# stages in graph clustering}
		\label{fig:layers}
	\end{subfigure}
	\caption{The influence of the parameters on the accuracy of GED approximation, best viewed in color} 
	\label{fig:parameter}
\end{figure}

In Figure~\ref{fig:parameter}, we check the influence of the hyper-parameters on the performance of HGMN. Figure~\ref{fig:vectors} shows that when the number of eigenvectors used for embedding pooling increases, the performance of HGMN first increases and then stabilizes. Recall that the number of eigenvectors decides the number of correlation matrices provided by each stage for the downstream CNN. As more eigenvectors are used for pooling, more information in the node embedding of the previous graph clustering stage is kept and thus more information is provided for the CNN. However, with a sufficient number of eigenvectors, adding new eigenvectors does not help as the first eigenvectors (correspond to the largest eigenvalues) already encode the most significant signals in the embedding matrix $\mathbf{H}^l_k$.

Figure~\ref{fig:layers} shows that when the number of graph clustering stages increases, the performance of HGMN also first increases but then saturates, similar to the case of pooling eigenvectors. This is because when using too many stages, each stage will only make a small change in the graph structure (e.g., groping two nodes into one) and thus does not provide too much information. For the PTC dataset, the performance of HGMN stabilizes with 5 stages; while for the AIDS dataset, the performance of HGMN stabilizes with 3 stages. This is because graphs in PTC are larger and can have more meaningful stages. This phenomenon also suggests that more stages are required for even larger graphs, on which HGMN can achieve even greater performance improvement than existing baselines as they do not use hierarchical clustering.       

\subsection{Efficiency Comparison}

\begin{figure}[!t]	
	\centering 
	\includegraphics[width=0.98\columnwidth]{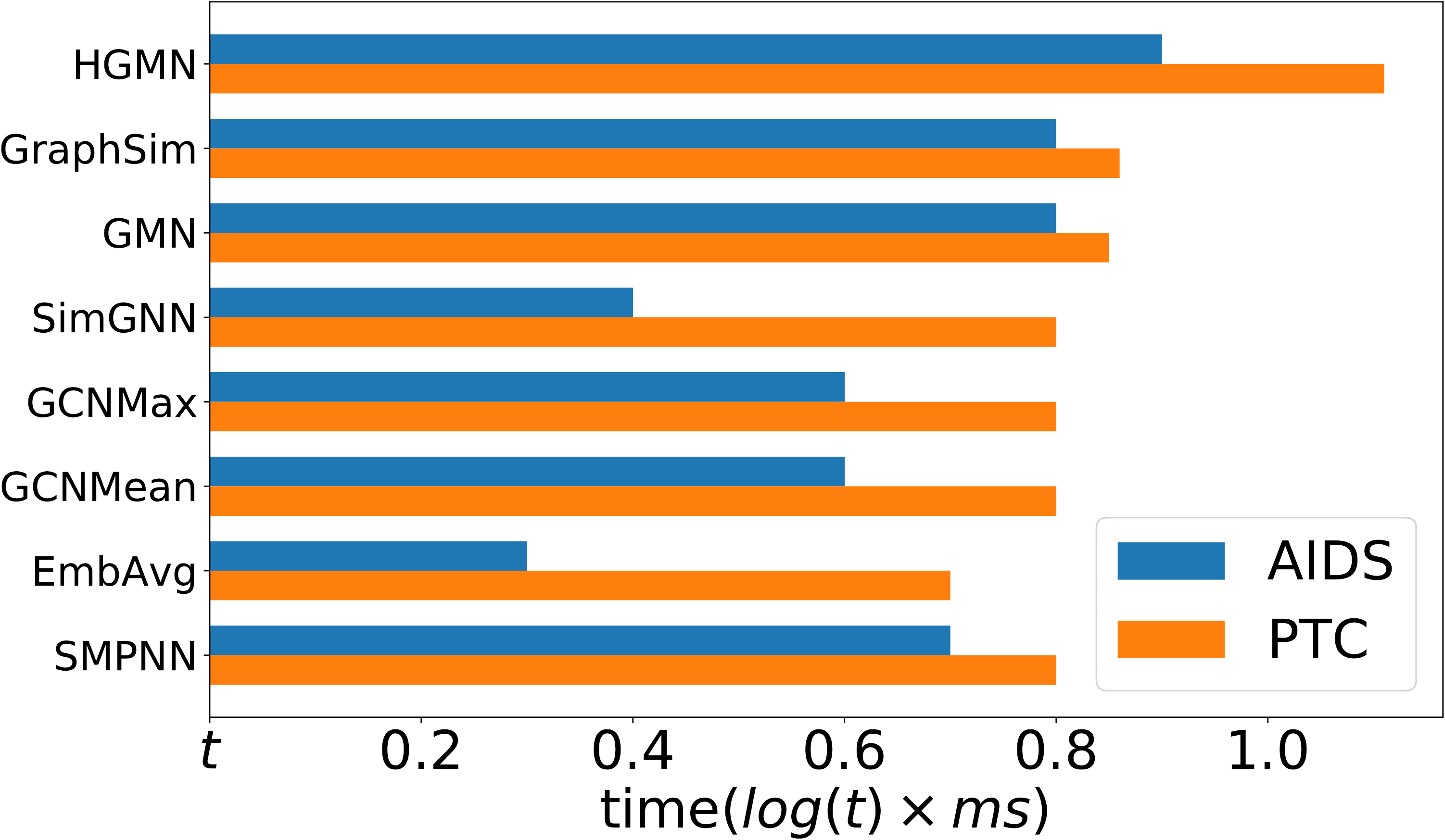} 
	\caption{The average GED computation time for some methods} 
	\label{fig:Efficiency}
\end{figure}

We report the average query processing time for GED similarity search for different methods in Figure~\ref{fig:Efficiency}. Query processing time is the time taken to compute the approximate GED between a query and all dataset items, and the reported results are measured on a machine with Intel(R) Xeon(R) E5-2697 v3 @ 2.6GHz CPU (56 physical cores) and 512GB RAM in single thread mode. We did not include the graph theory based methods (e.g., Beam~\cite{neuhaus2006fast}, Hungarian~\cite{riesen2009approximate} and VJ~\cite{fankhauser2011speeding}) as they are shown to be orders of magnitude slower than learning based methods~\cite{graphsim}. We used a dataset with small graphs (AIDS) and a dataset with large graphs (PTC) to check the influence of graph size. 

The results show that HGMN takes more time than the other methods because it uses hierarchical graph clustering and explicit node alignment. The higher computation complexity of HGMN is more obvious for larger dataset (PTC vs. AIDS) as larger graphs need more hierarchical clustering stages and make node alignment more complex. However, HGMN is not significantly slower than the other methods (e.g., 29.1\% and 12.5\% slower compared with GraphSim on PTC and AIDS, respectively) because graph neural network computation on the original graph dominates the overall complexity (required by all the methods). EmbAvg is the most efficient among all methods as it uses a simple dot product between the averaged embeddings of two graphs but its accuracy is poor according to Table~\ref{tab:GED} and Table~\ref{tab:MCS}. We think HGMN offers a reasonable trade-off between accuracy and efficiency by using a small increase in complexity to trade for better accuracy.

\section{Conclusions}

In this paper, we proposed the hierarchical graph matching network (HGMN) for efficient graph similarity computation. Motivated by the observation that two similar graphs should also be similar when they are clustered into more compact graphs, HGMN uses hierarchical clustering to provide the learning algorithm a multi-scale view of the differences between graphs. In addition, HGMN also adopts techniques including eigenvector based embedding pooling and earth mover based node alignment to build a complete machine learning pipeline. Experimental results on 4 datasets and 4 performance metrics show that HGMN consistently outperforms the baselines. Moreover, there are evidences that HGMN can scale to large graphs. 

%\input{abstract}
%\input{1_introduction}
%\input{2_related work}
%\input{3_method}
%\input{4_experiment}
%\input{5_conclusion}

%\bibliography{pyramid}
\bibliographystyle{unsrt}

\bibliography{HGMN}

\end{document}